# High carrier lifetimes in UMG multicrystalline wafers after P- diffusion compatible with high-efficiency cell structures

N. Dasilva-Villanueva[1,*], B. Arıkan[2], H. H. Canar[2,3], D. Fuertes Marrón[1], B. Hong[1], A. E. Keçeci[2,4], S. K. Bütüner[2], G. Bektaş[2,4], R. Turan[2,3], C. del Cañizo[1]

[1] Instituto de Energía Solar, Universidad Politécnica de Madrid, ETSI Telecomunicación, Av. Complutense 30, E-28040 Madrid (Spain)

[2] Center for Solar Energy Research and Applications (ODTÜ-GÜNAM), Ankara, Turkey

[3] Department of Physics, Middle East Technical University, Ankara, Turkey

[4] Micro and Nanotechnology, Middle East Technical University, Ankara, Turkey

**Abstract**

High-quality multicrystalline Upgraded Metallurgical Grade Silicon (UMG-Si) offers significant advantages over conventional polysilicon-based PV technology, associated to lower cost, lower energy budget and lower carbon footprint. The aim of this study is twofold: on the one hand, to ascertain the efficiency potential of solar cells based on this material in terms of carrier lifetime; and on the other hand, to explore, as a result of that, the adoption of high-efficiency cell architectures by establishing an effective rear-side passivation scheme for the implementation of passivated emitter rear contact (PERC) devices. The carrier lifetime and the surface passivation efficacy are investigated for different passivating layer configurations after single and double P-diffusion gettering processes. Layer stacks consisting of $Al_2O_3$, $SiO_xN_y$ and a-$SiN_x$:H capping overlayers have been optimized, on industrial size, saw-damage-etched UMG wafers and results compared to those obtained using reference iodine-ethanol (IE) passivation. Diagnosis based on minority carrier lifetime and implied $V_{oc}$ ($iV_{oc}$) measurements helped monitor the impact of parameter optimization on wafer quality, particularly after firing processes. Carrier lifetimes over 600 µs at $\Delta n=10^{15}$ cm$^{-3}$ injection level as well as up to 790 µs locally have been measured in UMG-Si wafers passivated with IE after a Phosphorus Diffusion Gettering (PDG), demonstrating the suitability of the material for high-efficiency cell architectures. Values higher than 300 µs have been obtained with $Al_2O_3$-based passivation layers for gettered UMG wafers, with implied $V_{oc}$ values up to 710 mV. These record-breaking lifetimes and $iV_{oc}$ figures obtained with p-type multicrystalline UMG-Si material demonstrate a significant upgrading of its electronic quality by means of industry-scalable technical processes.

[†]Corresponding author: nerea.dasilva@ies.upm.es

# 1. Introduction

The Paris Agreement established both a net-zero emissions milestone by 2050 as well as a ceiling limit of 2.0ºC raise in the global temperature in order to successfully prevent severe consequences related to climate change [1]. To accomplish it, global greenhouse gas (GHG) emissions need to be rapidly reduced from the current 43 billion tons of $CO_2$eq emitted annually worldwide. With the energy sector accounting for 73.2% of the global emissions [2], achieving a decarbonized system is particularly important, as fossil fuels still have a predominant presence, with the total electricity production being still fossil-based as of 2020 [3].

In this scenario, solar photovoltaics (PV) is called to play a main role in the coming years. As its levelized cost of electricity (LCOE) has been proven lower than any other electricity source in many parts of the world [4], and with annual additions to the net capacity surpassing the hundred GW for a cumulative 707 GW in 2020 [5], it is evident that PV will be the backbone of the future decarbonized energy system [6]. Today, crystalline silicon represents around 95% of all the PV production technology [7], and while the technology has got increasingly cheaper throughout the years, spot market prices are subject to fluctuations, with recent spikes up to a 300% increase [8,9], being especially notable in Siemens purified polysilicon.

Upgraded Metallurgical Silicon, UMG-Si, results from a different purification technique and has been proposed as an alternative feedstock for solar cells. It provides not only lower production cost and lower CAPEX, but also a manufacturing process with lower environmental impact, with an overall 20% reduction of GHG emissions, according to recent estimations [10]. UMG-Si typically presents a higher content of impurities, such as transition metals like iron, copper, and nickel, as compared to conventional polysilicon. Additionally, crystallographic defects might appear during growth because of the presence of these contaminants in the feedstock [11]. Furthermore, the simultaneous presence of B and P in the melt causes a moderate compensated character in the resulting ingots that can be further tuned by the deliberate addition of dopants during the crystal growth, and whose impact in carrier trapping and recombination needs to be evaluated [12–14]. Despite all these issues, multicrystalline UMG-Si solar cells have already demonstrated a maximum conversion efficiency of 20.76% [15] in a commercial production line, at the same level as that obtained in the same line using conventional poly-Si.

The objective of this work is the improvement of the electronic quality of multicrystalline p-type UMG-Si wafers by means of finely tuned processing steps, allowing the manufacture of high-efficiency devices and thereby fully exploiting the cost and environmental advantages of UMG. In order to effectively remove impurities from the bulk of the material, phosphorous diffusion gettering (PDG) steps are introduced under different conditions prior cell processing. As the P-diffusion conditions that lead to the largest improvement observed in carrier bulk lifetime do not correspond to those required for an optimal P-emitter profile, particularly with respect to carrier recombination, the proposed strategy is the realization of a pre-gettering step optimized for defect annihilation and followed by an etch-back, before the realization of a second P-diffusion for P-emitter provision. This strategy has been successfully implemented, leading to record values of carrier lifetime in this material.

Scalable passivation schemes have also been tested, based on the combination of alumina ($Al_2O_3$) and hydrogenated silicon nitride (a-$SiN_x$:H) dielectric layers, which is widely used for the rear side passivation of passivated emitter rear contact (PERC) solar cells, due to the low interface trap density ($D_{it}$) and high density of negative fixed charges (on the order of $10^{12}$ cm$^{-2}$) of the former, allowing both chemical- and field-effect-based surface passivation [14], and the diffusion of atomic hydrogen from the latter, passivating surface and bulk defects [16].

Additionally, this passivating structure benefits from the fact that hydrogen-rich dielectric layers release hydrogen that diffuses into the bulk during the firing steps required for metallisation. The diffused atomic hydrogen interacts with impurities and defects present at the interface between wafer and coating, such as dangling bonds, and passivate them. Thus, a high hydrogen content in the passivation layer contributes significantly to the passivation quality.

## 2. Experimental methods

The UMG-Si material used in this study was manufactured via the metallurgical route by FerroGlobe, and subsequently crystallized as multicrystalline ingots by FerroSolar [15]. UMG compensated p-type silicon wafers were obtained thereof, with standard dimensions $156.75 \times 156.75$ mm$^2$, $200 \pm 20$ μm thickness and resistivity ranging between 1.15 and 1.3 Ω·cm. The compensation level defined as $(N_A+N_D)/(N_A-N_D)$ where $N_A$ and $N_D$ are the concentrations of p-type and n-type dopants, respectively, varies along the ingot between 2 and 4 [17].

### 2.1. PDG optimization studies

The wafers used in the analysis of the PDG process were firstly characterized by Photoconductance Decay (PCD) measurements, both before and after chemical passivation with iodine-ethanol (IE), and then chemically treated with CP4 ($HNO_3$/HF) etching, RCA1 ($HN_3$:$H_2O_2$) surface cleaning and a treatment in 2% HF to ensure native oxide removal, before the thermal process. P-diffusion was performed in a tubular furnace in $O_2$ and $N_2$ atmosphere, at both low (780ºC) and high (850ºC) temperatures and varying the processing time, followed by a 10-minute drive-in. The P-source was liquid $POCl_3$ kept at 26ºC, assuring super-saturation. After PDG, the emitter formed atop the surface was removed by CP4 etching and PCD measurements were again conducted before and after IE passivation.

To evaluate the possibility and the magnitude of further lifetime improvements, a second PDG and subsequent etching and characterization were carried out following the same steps.

Complementary PDG experiments have been carried out in an industrial-type pilot line, providing further information on the conditions under which PDG can be carried out. Samples that underwent both the PDG and the P-diffusion emitter provision were etched by a two-sided saw damage etching (SDE) with 5% KOH solution for 4 minutes at 75ºC and cleaned afterwards with ozone and acidic baths. Prior to each deposition step, the native oxide was removed in 5% HF solution. The PDG plus P-emitter process was carried out in an atmospheric pressure diffusion furnace in three stages: a pre-deposition phase for 10 minutes at 780ºC, $POCl_3$ deposition for 12 minutes at 830ºC, and 7 minutes drive-in at 830ºC. After the gettering process, both surfaces are etched using the RENA Inpilot industrial single side etching (SSE) tool in order to remove from either side the doped regions prior to $Al_2O_3$ deposition.

## 2.2. Passivation studies

Figure 1a shows the different structures that were tested in the passivation studies. Firstly, samples with a 120 nm thick layer of silicon nitride ($SiN_x$:H) were tested and used as a reference to evaluate the impact of hydrogen diffusion across alumina barriers in subsequent experiments, as explained later. Then, to create field-effect passivation by means of band-bending, a thin layer of alumina, $Al_2O_3$, was introduced. This field-effect passivation is greatly effective in p-type silicon, but if the layer thickness is too large, the chemical passivation induced by the $SiN_x$ layer would be greatly reduced. The idea of these studies is to find a compromise between chemical and field-effect passivation in order to maximize the overall passivation, using the alumina layer thickness as a free parameter. A third option, using silicon oxynitride ($SiO_xN_y$:H) instead of $Al_2O_3$ as an intermediate phase between the native silicon oxide and silicon nitride, was also tested.

The $Al_2O_3$ passivation film was deposited on both sides of the wafer using atomic layer deposition (ALD, SolayTec). Trimethylaluminum (TMA, $Al_2(CH_3)_6$) and water vapor were used as precursors and the deposition was carried out for 37, 74 and 111 cycles, with calculated deposition thicknesses of about 5, 10 and 15 nm, respectively. Subsequent annealing (activation) under $N_2$ was performed at 425°C for 15 minutes. The passivation layer stack was completed with a 120 nm thick silicon nitride ($SiN_x$) film, formed by two consecutive layers of 27 nm with a refractive index of 2.18 and 93 nm with a refractive index of 1.99, as capping layer. $SiN_x$ and hydrogenated amorphous silicon oxynitride ($SiO_xN_y$:H) layers were deposited using plasma-enhanced chemical vapor deposition (PECVD) in a **DongGuan** Plasma system. Afterwards, firing was applied in an industrial conveyor belt furnace (BTU International Inc.) at a peak temperature of approximately 735°C with a 380 cm/s belt speed. The full processing sequence is represented in

Figure 1b.

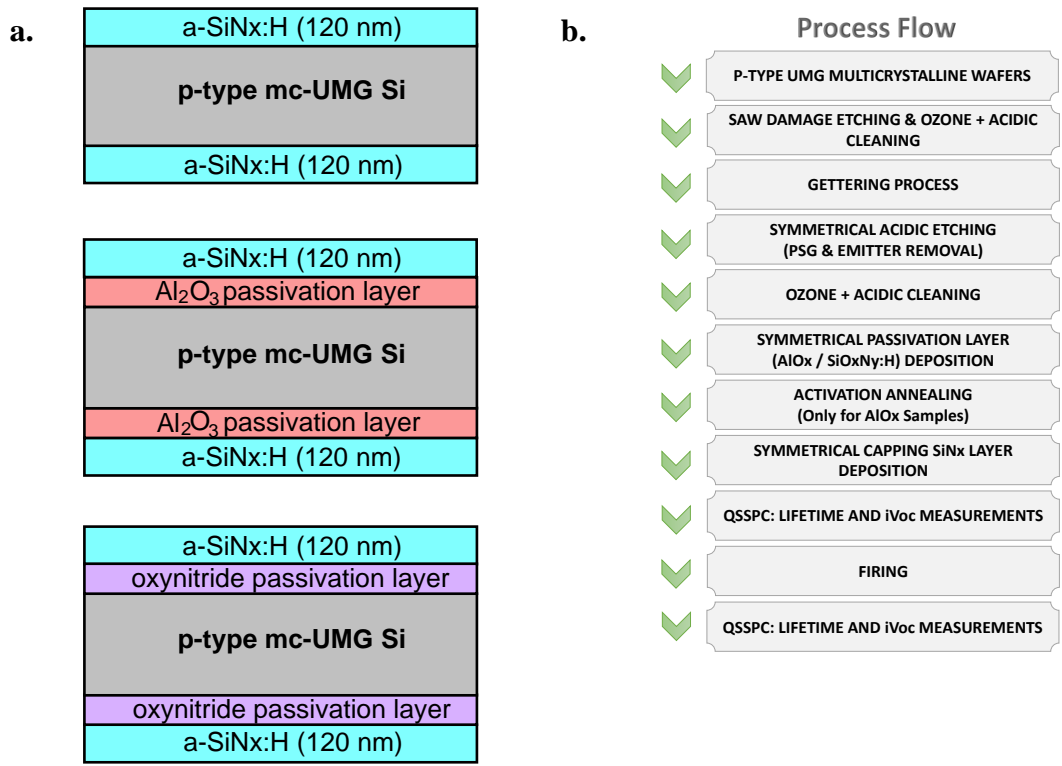

Figure 1. a) Passivation test structures. b) Processing sequence.

## 2.3. Measurements

PCD-based minority carrier lifetime measurements were done with Sinton instruments WCT-120 and WCT-120TS in transient mode, with a flash characteristic decay time of 23 µs as measured at the reference cell. For the acquisition of the lifetime curves, an optical constant value of 0.7 was used, along with Dannhäuser's carrier mobility model. Although it is accepted that alternative mobility models such as Klaassen's [18] or Schindler's [19] are generally more appropriate for compensated semiconductors, it turns out that, for the compensation and the carrier injection levels achieved in our material, differences among mobility models are not relevant, as shown in previous works [20]. Photoluminescence (PL) mapping has been carried out with the Semilab PLI 1001 setup, using an IR-laser in a focused stripe over a moving wafer. The setup is capable, through in-situ calibration, to convert PL images into carrier lifetime maps.

## 3. Results and Discussion

### 3.1. Gettering processes

Effective lifetimes measured by PCD at an injection level of $10^{15}$ cm$^{-3}$ are shown in Figure 2 for several representative samples subjected to different processing conditions.

In order to obtain good statistical significance, wafers are cut in 5x5 cm$^2$ samples, and 96 samples from 12 wafers sliced from a single brick were analyzed in 12 separate batches. First PDGs are carried out at 780ºC and 60 minutes, as this process was previously optimized for UMG silicon [21], and second PDGs were carried out at different combinations of time and temperature, in order to analyze the impact of processing conditions on carrier lifetime improvement. Dashed lines in Figure 2 are a guide to the eye to help follow each sample's path.

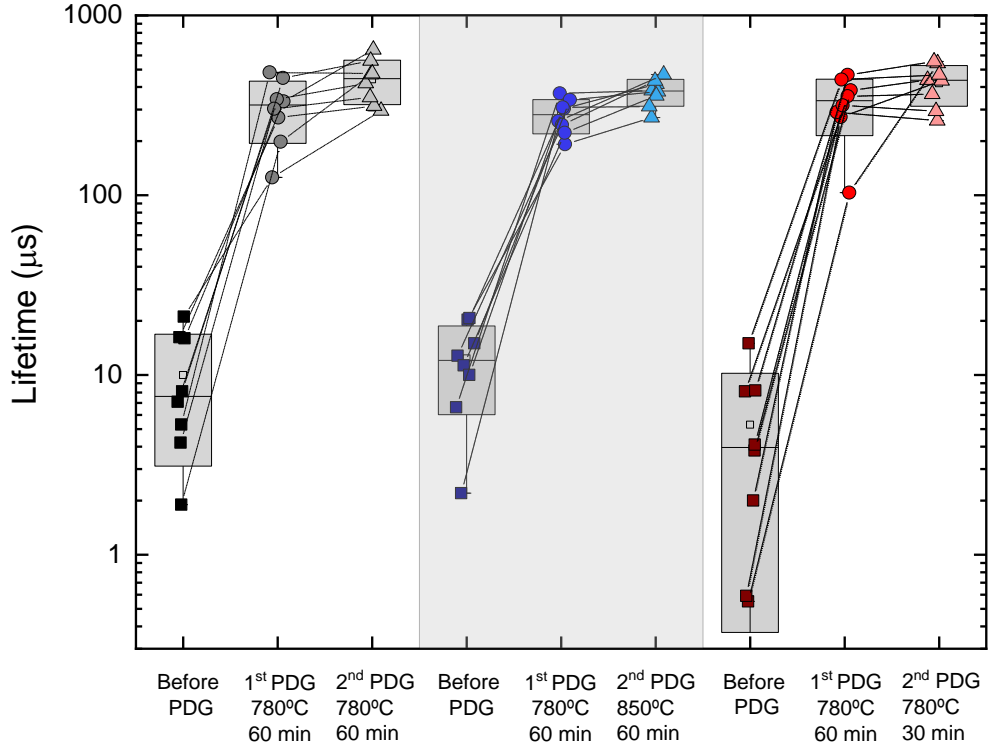

Figure 2. Effective lifetime variation after one and two PDG processes for different conditions of iodine-ethanol-passivated UMG-Si wafers. In the box-type plot whiskers represent max-min variation, the height of the box represents the standard deviation, the non-filled square corresponds to the mean value and the middle line represents the median value. Connection dashed lines are included to identify the change of individual samples.

A substantial improvement is shown after the first PDG, with values ranging from 10 up to almost 500 µs, i.e., a 50x improvement ratio in most cases. The second PDG results in further but moderate lifetime improvements, as shown by the right boxes in each panel. Maximum improvements up to 5x were recorded with respect to results from the first PDG. Overall, the compound effect of the two PDG processes results in total lifetime improvements that range between 20 to almost 1000 times the starting effective lifetimes.

Interestingly, the fact that the second PDG can be carried out at higher temperatures than the first PDG, as it can be seen in Figure 2, opens the possibility of taking advantage of the gettering action during the P-diffusion process for emitter formation. In this manner, a single additional P-diffusion process and the subsequent chemical removal of the surface regions, to be included in the processing line before the actual emitter provision, can significantly upgrade the electronic properties of UMG-Si wafers and thereby the performance of UMG-Si-based solar cells.

These results demonstrate the feasibility of reaching average lifetimes over 300 µs in p-type, multicrystalline UMG-Si. This is an outstanding result, which opens the possibility of implementing high-efficiency cell architectures using UMG-Si. According to the analysis carried out by Hofstetter et al. [22], to obtain efficiencies higher than 21% in PERC architectures, threshold carrier lifetimes around 200 µs must be ensured at maximum-power-point (MPP) operation conditions. As a matter of fact, the MPP in devices operating under standard AM1.5G spectrum typically occurs at an injection level of photogenerated excess carriers below $10^{15}$ cm$^{-3}$.

As it can be seen in Figure 3, lifetime measurements in transient conditions for the studied samples reveal curves that consistently show higher lifetimes at $\Delta n=10^{14}$ cm$^{-3}$, despite an increased noise level, than those observed at $10^{15}$ cm$^{-3}$. In this respect, values referred at $\Delta n=10^{15}$ cm$^{-3}$ should be considered as conservative estimations, likely underestimating the actual carrier lifetimes at MPP.

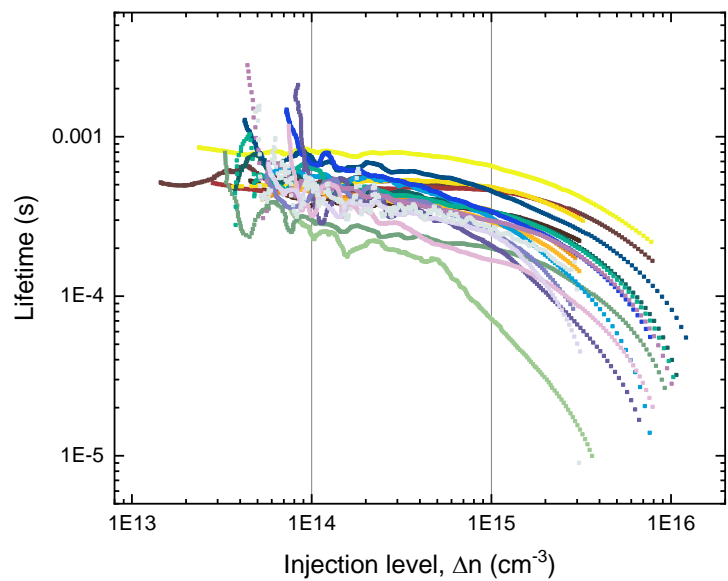

Figure 3. Assortment of carrier lifetime curves measured in transient mode. Vertical lines are shown at $10^{14}$ cm$^{-3}$, typical values for the MPP in PERC structures, and $10^{15}$ cm$^{-3}$, less affected by noise, to which values in this study are referred.

## 3.2. Passivation studies

In this section, we study the carrier lifetime values (measured at an injection level of $10^{15}$ cm$^{-3}$) in UMG-Si samples resulting from different passivation-coating strategies, with special attention to the impact of the firing process associated to subsequent screen-printing metallization of complete solar cells. Figure 4a shows experimental results obtained from the different passivation architectures considered, both before and after the firing process as described in the Experimental section. A carrier lifetime map obtained after firing from PL measurements of a sample processed with the best performing passivation structure (5 nm $Al_2O_3$/ 120 nm $SiN_x$ layer stack) is also shown in Figure 4b in order to determine spatially resolved variations in lifetime across the sample. Such combination of experimental results will be used to compare the electronic behavior of samples before the PDG, after one PDG and after two PDGs, following the recipe previously indicated for the pilot line.

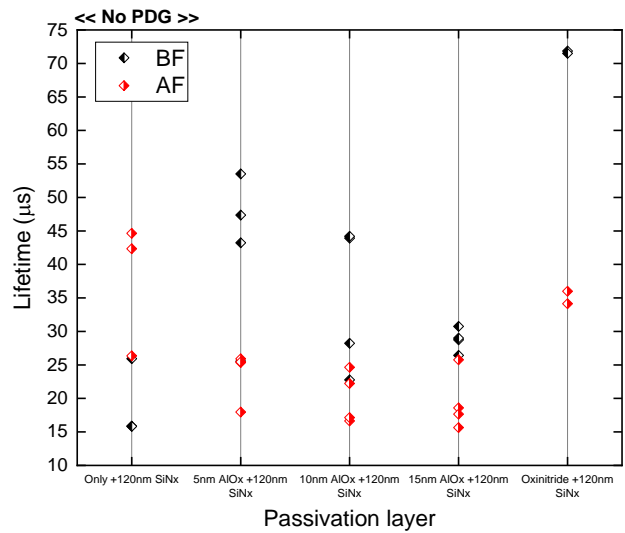

(a)

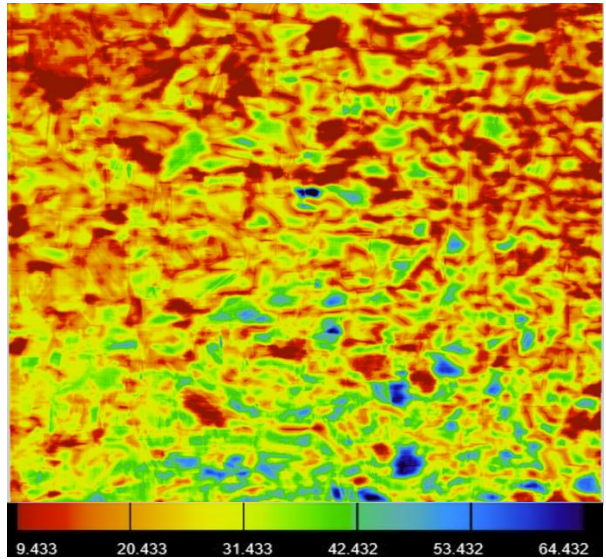

(b)

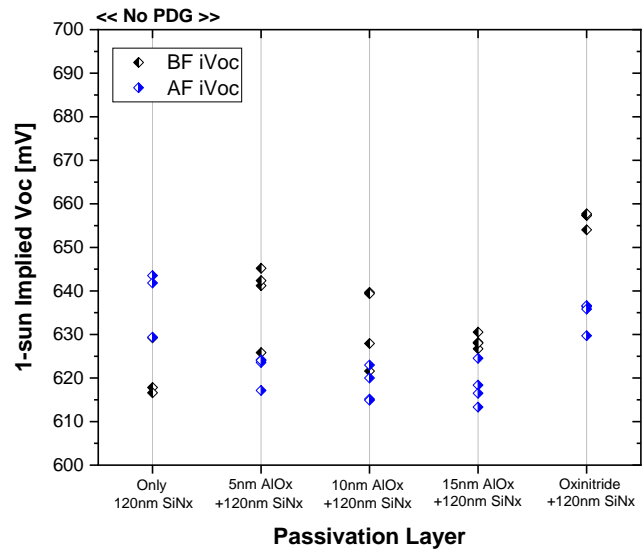

(c)

Figure 4

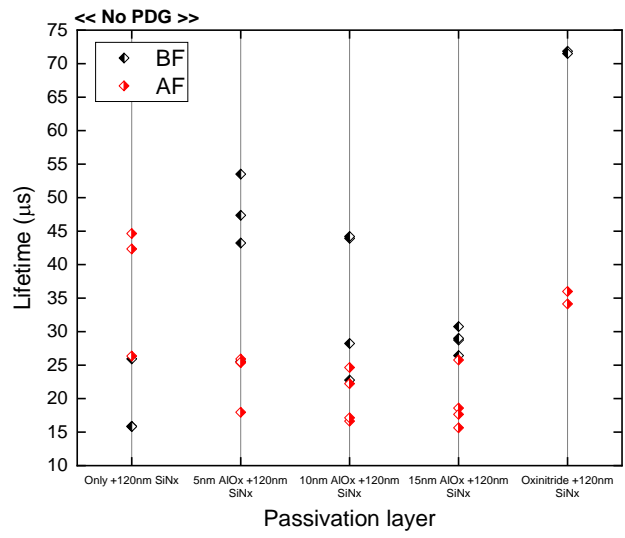

(a)

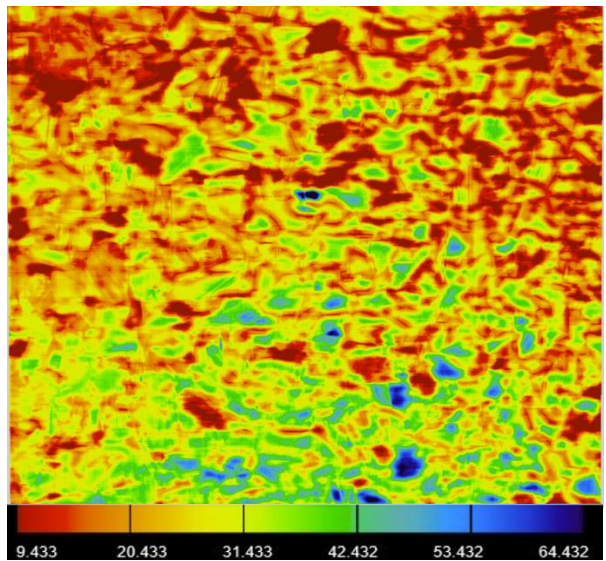

(b)

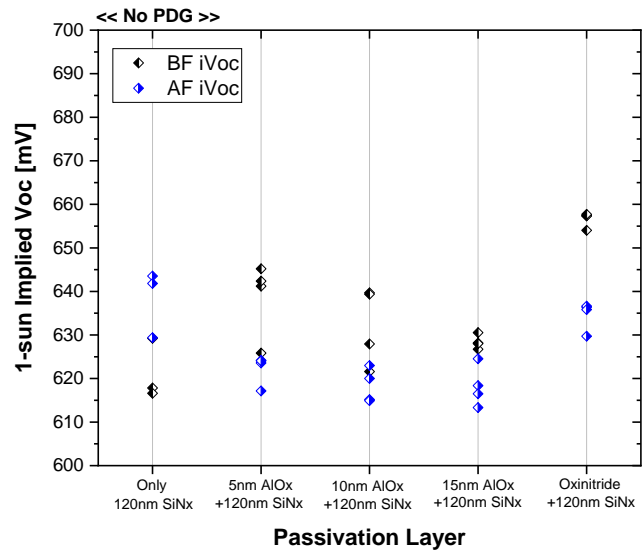

(c)

Figure 4. a) Carrier lifetimes obtained for different passivation structures in samples before P-diffusion, before and after firing. b) PL map after passivation with a stack of 5 nm $AlO_x$/120 nm $SiN_x$, showing average lifetime of 25.3 µs. c) Implied Voc values for different passivation structures of samples before P-diffusion.

Considering the pre- and post-firing measurement results, a decreasing trend is observed in the samples with alumina layer after firing, while the reverse trend is observed in samples provided just with a $SiN_x$ layer. This behavior is likely due to the relatively high hydrogen transfer originating from the hydrogenated $SiN_x$ layer into the silicon wafer at the firing processing temperature, in absence of the alumina layer. Although as-deposited $SiO_xN_y$/$SiN_x$ stacks give relatively high lifetime results (up to 71 µs), these values are reduced after firing. $SiO_xN_y$ films seem to be advantageous in terms of field-effect passivation on p-type surfaces, as they show a lower positive fixed charge density compared to $SiN_x$ films [23].

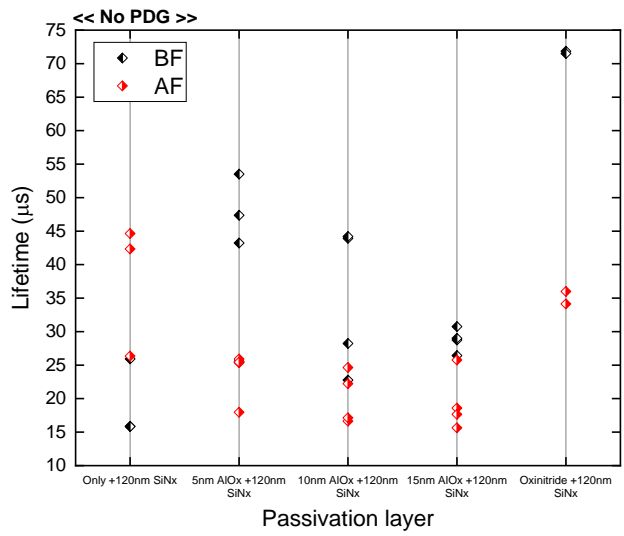

(a)

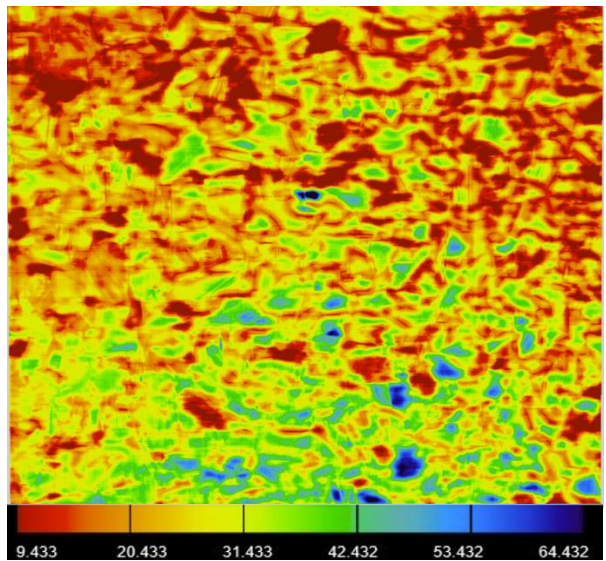

(b)

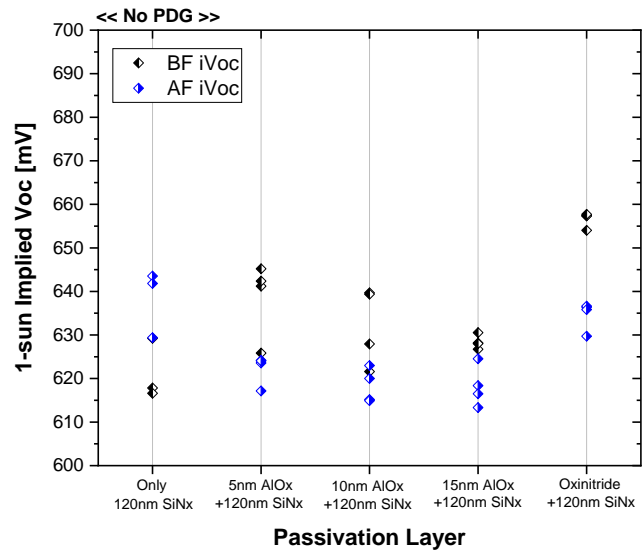

(c)

Figure 4b shows a PL mapping of a full wafer (15.6x15.6 cm$^2$), showing average carrier lifetime values around 25 µs. Several high- and low-valued areas can be observed, varying with grain distribution. While local maximum lifetimes of around 60 µs can be observed, the average lifetime within the grains is around 40 µs, greatly reduced to around 10 µs along the grain boundaries.

Figure 4c shows iVoc results for different passivation structures. Similar values are recorded for all types of passivation, with the highest ones obtained before firing from samples passivated with oxynitride.

Figure 5 shows carrier lifetime values measured at an injection level of 10$^{15}$ cm$^{-3}$ for all the passivation structures studied, as well as the PL map of a representative sample measured after the first phosphorus diffusion gettering and subsequent emitter etching, provided with 5 nm Al$_2$O$_3$/ 120 nm SiN$_x$ layer stack passivation.

The improvement in the sample group that was treated with a single PDG treatment is quite remarkable, especially for samples provided with Al$_2$O$_3$/SiN$_x$. The improvement is somehow modest for the SiN$_x$ and oxynitride cases. As it can be seen, best results are obtained in all cases for a 5 nm thick alumina layer, reaching high carrier lifetime values up to 309 µs. The gradual worsening of lifetime values with the thickening of the alumina can be explained by a progressively inefficient chemical passivation resulting from a reduced hydrogen transfer from the $SiN_x$, with alumina acting as a barrier layer for hydrogen diffusion.

For samples only passivated with hydrogenated compounds, the recorded carrier lifetime improvement is greatly increased after firing, which triggers hydrogen diffusion from the coating directly into the wafer. For samples passivated by alumina and silicon nitride layer stacks, the firing process does not seem to have much effect at the light of the variable results obtained for the three different alumina thicknesses studied. In such cases, field-effect passivation induced by the presence of the alumina layers seems to clearly outperform that of a chemical origin related to hydrogen diffusion.

(a)

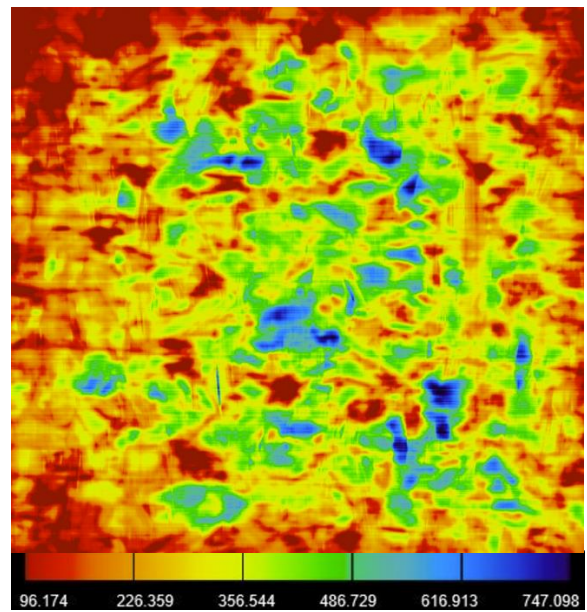

(b)

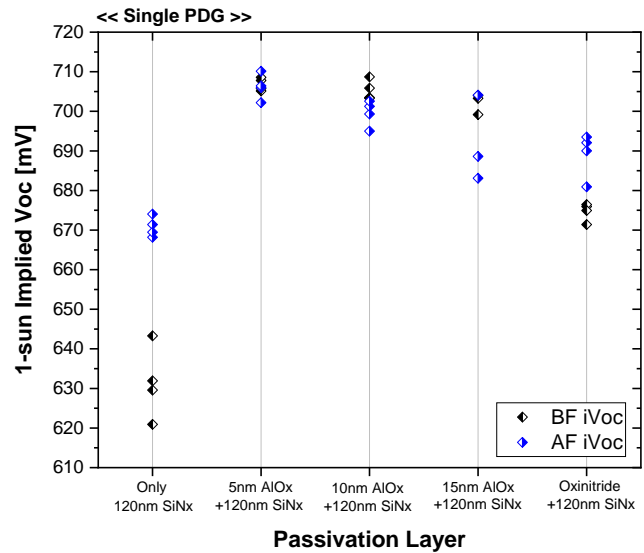

(c)

Figure 5b shows a PL-based carrier lifetime map of a representative PDG-treated sample, after etching and passivation, showing overall one order of magnitude improvement throughout the wafer. Notable differences between grains are still observed, with local maxima clearly surpassing 700 μs. The average carrier lifetime is 300 μs, approximately 12 times the average lifetime of a non-gettered sample.

iVoc values are represented in Figure 5. Maximum values are obtained for a 5nm $Al_2O_3$/$SiN_x$ stack, reaching up to 710 mV. This result is remarkable in comparison to other studies carried out in PERC solar cells [24,25], proving excellent passivation and emphasizing potential for high solar cell efficiencies.

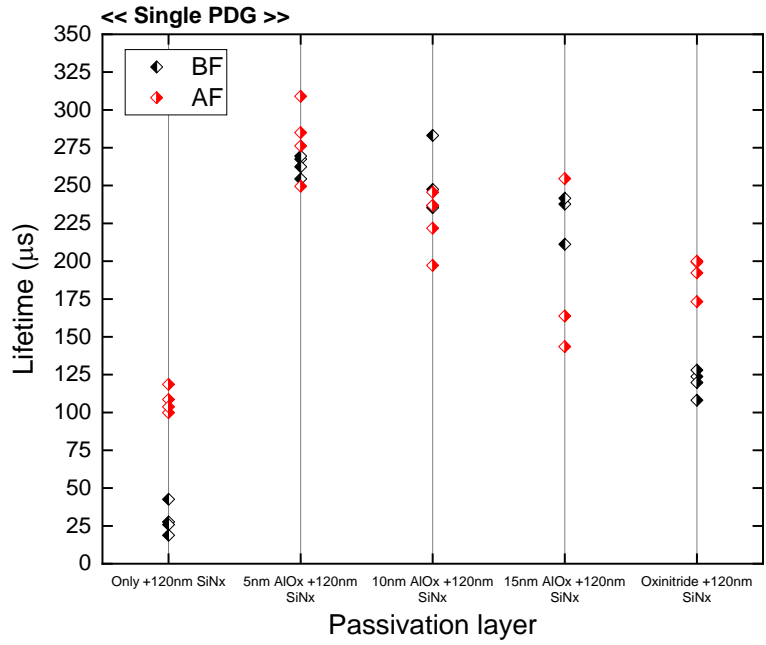

(a)

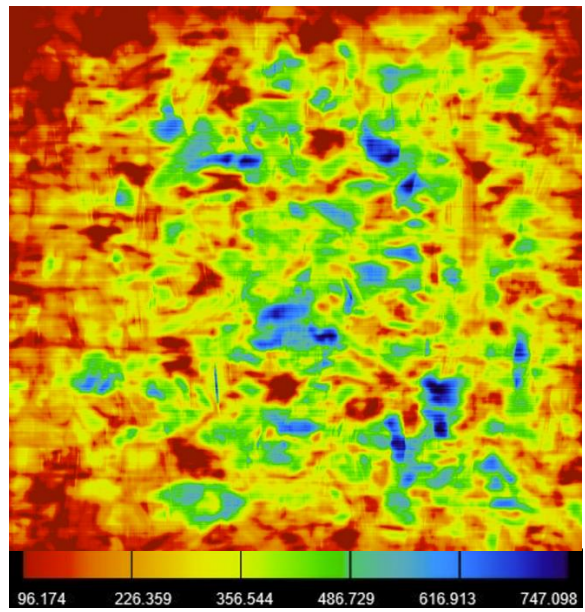

(b)

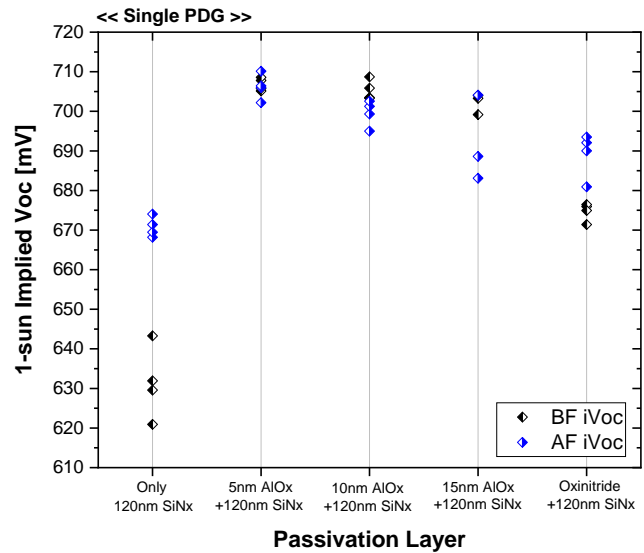

(c)

Figure 5. a) Carrier lifetimes obtained for different passivation structures from samples after one P-diffusion gettering process. b) PL-based carrier lifetime map of a representative sample after one PDG and etching, passivated with a stack of 5nm AlOx/120nm SiNx, showing an average lifetime of 300.2 μs. c) iVoc values for different passivation structures of samples after one PDG process.

Finally, Figure 6 shows similar figures for the case of samples that underwent a second P-

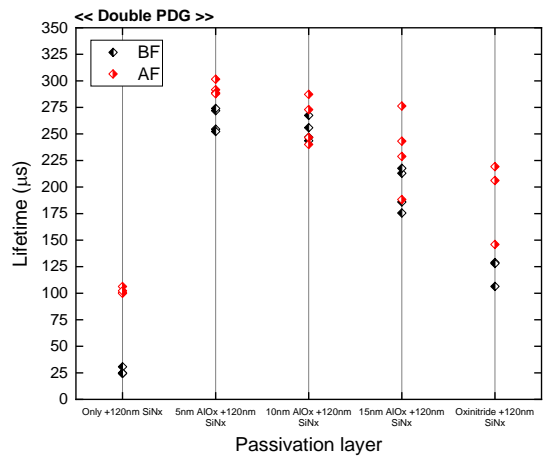

diffusion gettering process. In

(a)

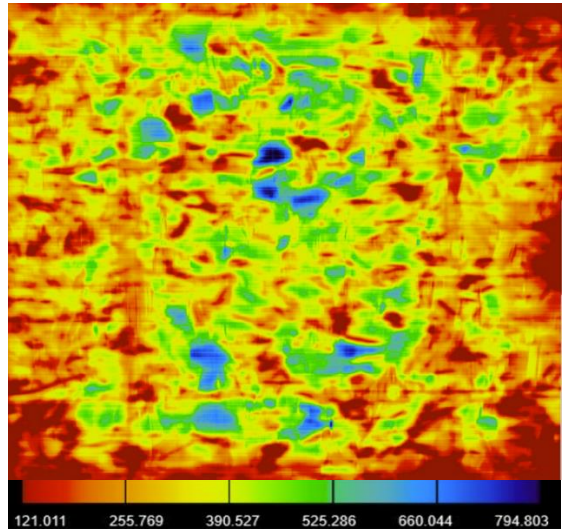

(b)

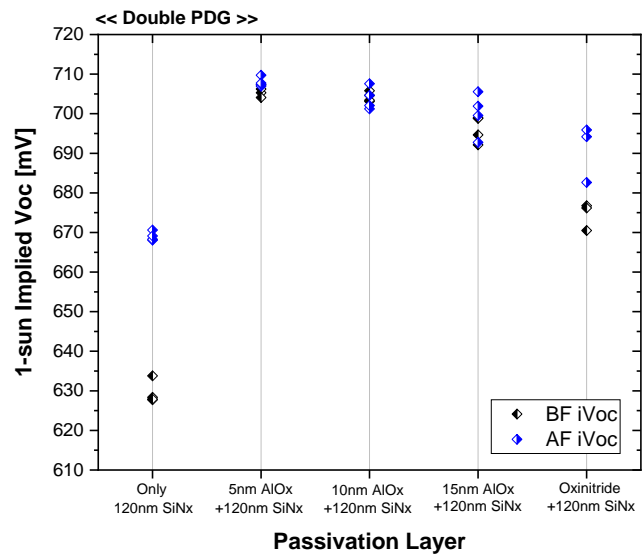

(c)

Figure 6

(a)

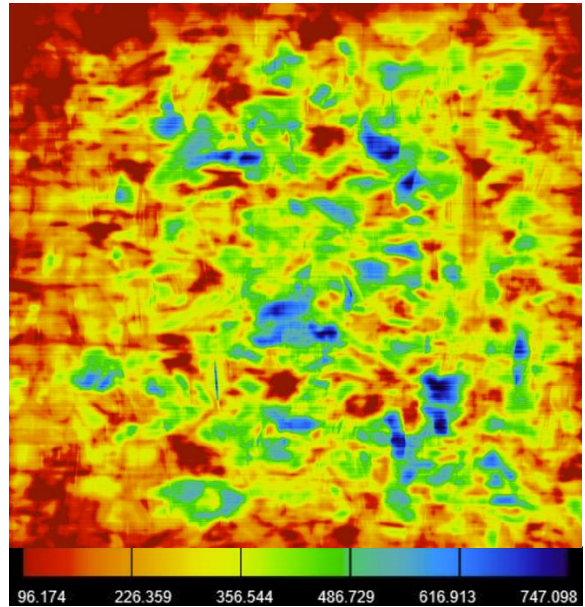

(b)

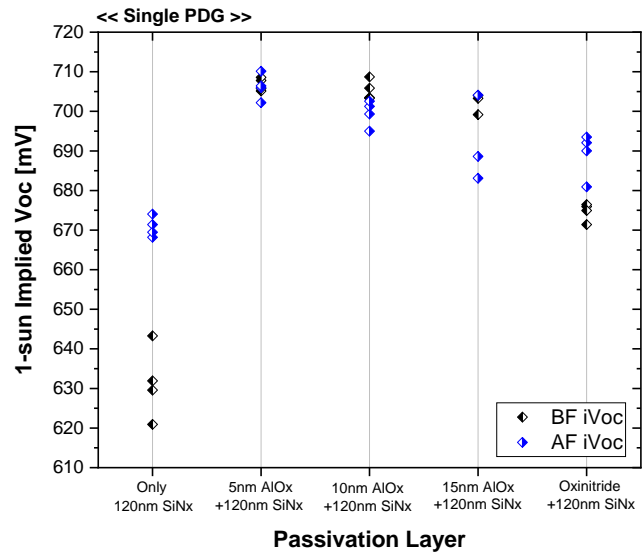

(c)

Figure 5a no significant effect of the second PDG treatment is observed in the sample group that was only treated with $SiN_x$ passivation. In the $Al_2O_3/SiN_x$ passivation sample group, the general trend observed in the single PDG sample group as a function of the alumina layer thickness was also observed in doubly PDG-treated samples. Although the number of samples in each group is small, the dispersion observed in carrier lifetime values among different samples seems to decrease after the second PDG treatment, a tendency also observed in samples passivated with iodine-ethanol (see Figure 2). In addition, the increase in carrier lifetime values after the second PDG treatment in the sample group with 10 and 15 nm thick $Al_2O_3$ is remarkable. After the second PDG process, a further decrease in the density of impurity-induced defects in the silicon wafer

(a)

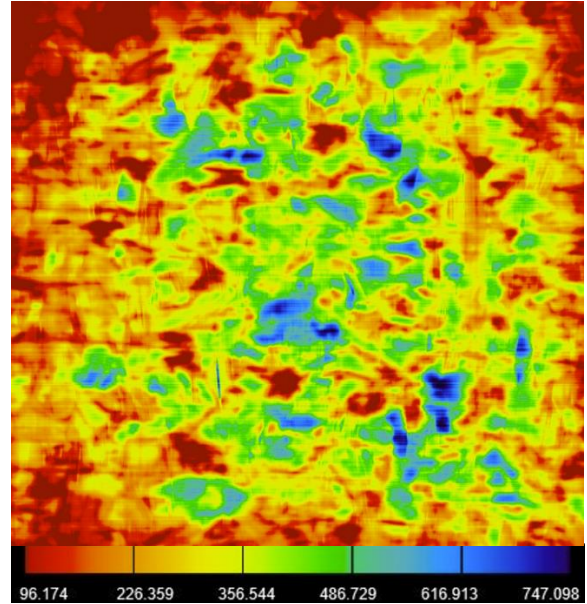

(b)

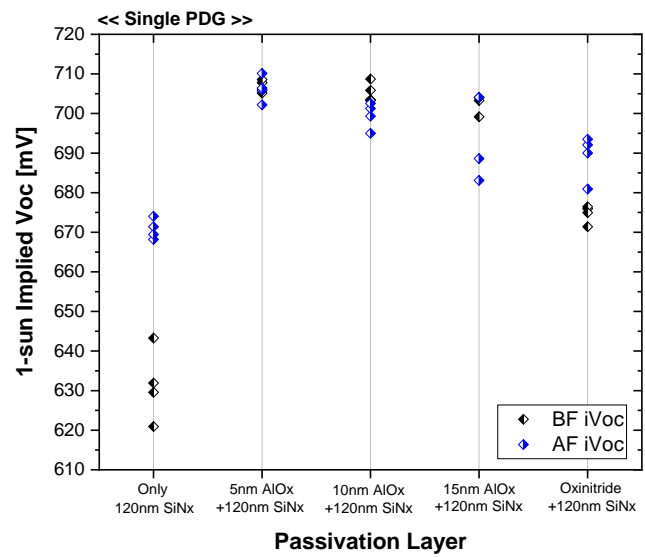

(c)

Figure 5 is expected. Again, the highest lifetime value was obtained with the passivation layer consisting of 5 nm $Al_2O_3$ thickness, as in the single PDG sample group (301 μs). In the sample group in which oxynitride was used, no significant effect is observed due to the second PDG.

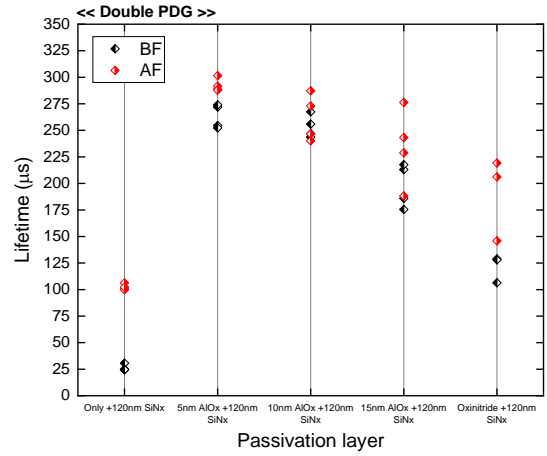

(a)

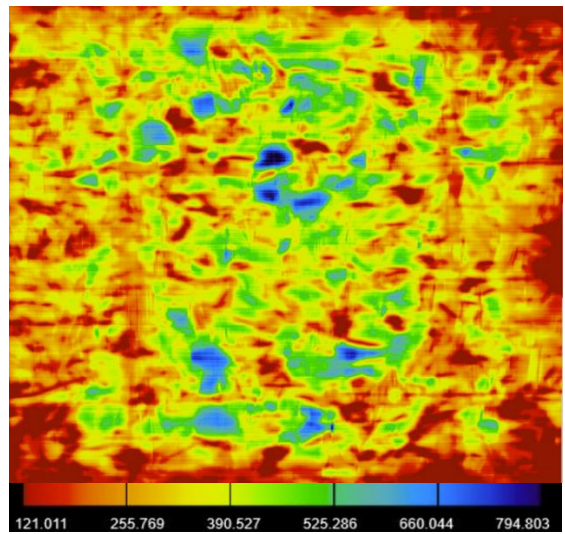

(b)

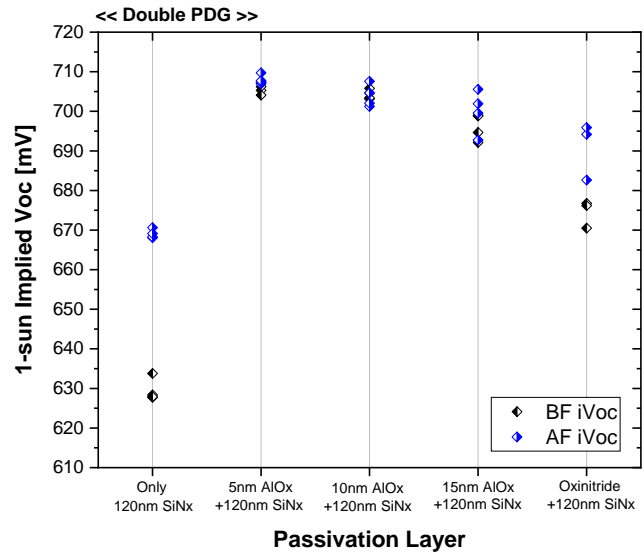

(c)

Figure 6b shows a PL map of a representative wafer after two subsequent PDG processes, after etching and with 5 nm $Al_2O_3$/120 nm $SiN_x$ layer stack passivation, showing local maxima as high as 790

(a)

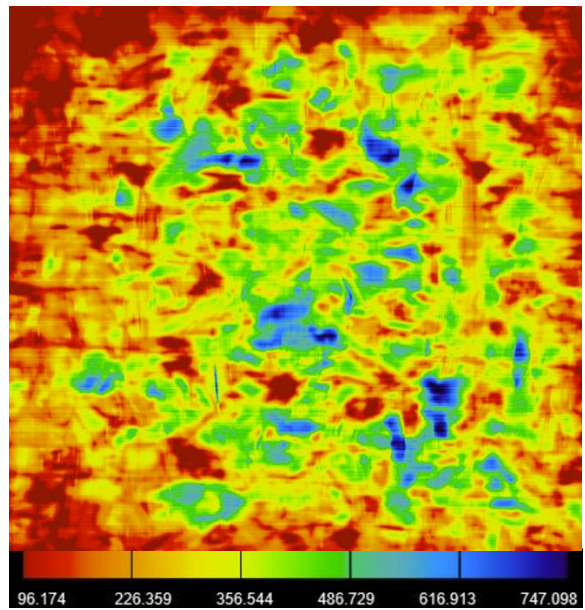

(b)

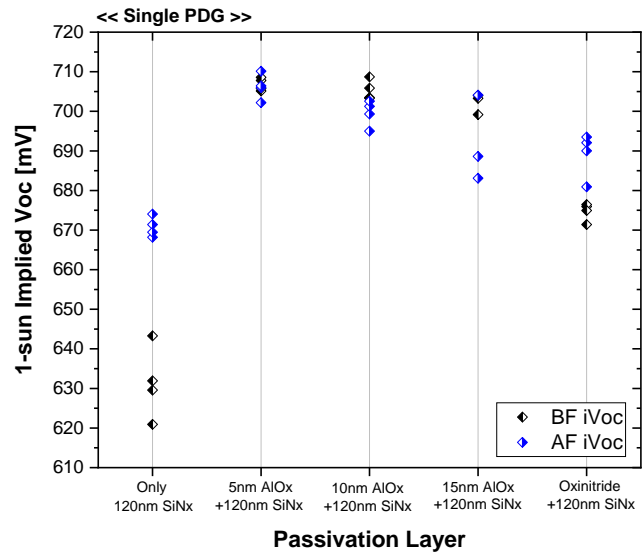

(c)

Figure 5

(a)

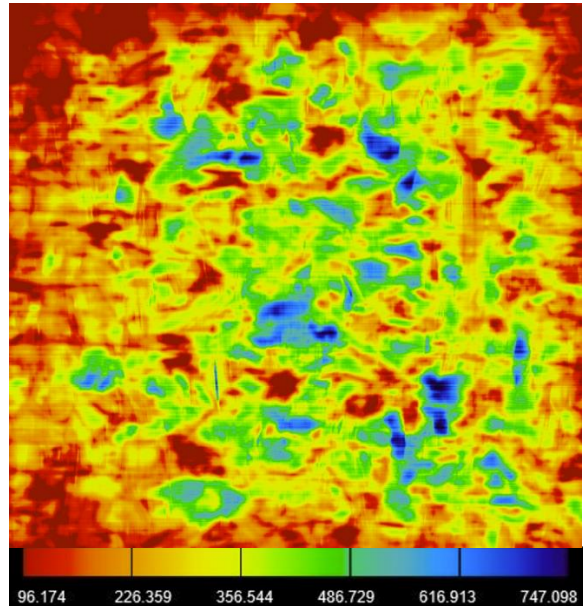

(b)

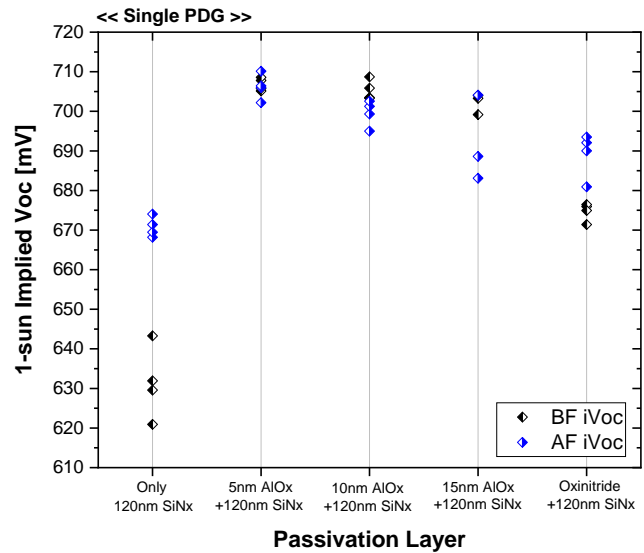

(c)

Figure 5μs, with an average lifetime of 325 μs. These results represent a moderate but still significant 10% carrier lifetime improvement with respect to the single PDG-processing of UMG-Si samples.

In Figure 6c, it can be seen how iVoc values show little to no further improvement in comparison to values obtained after a single PDG process.

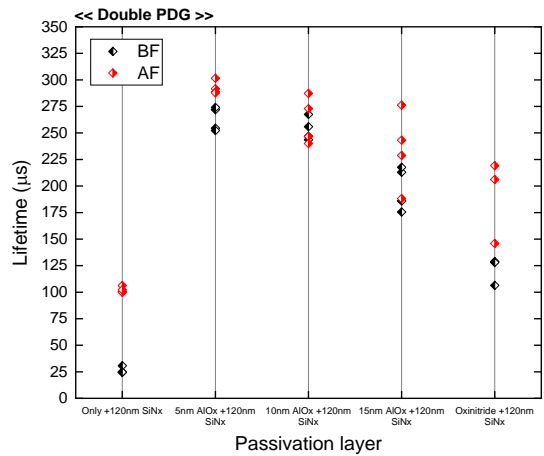

(a)

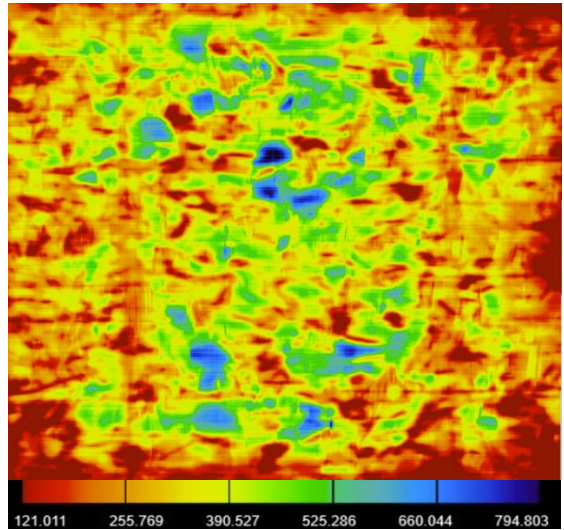

(b)

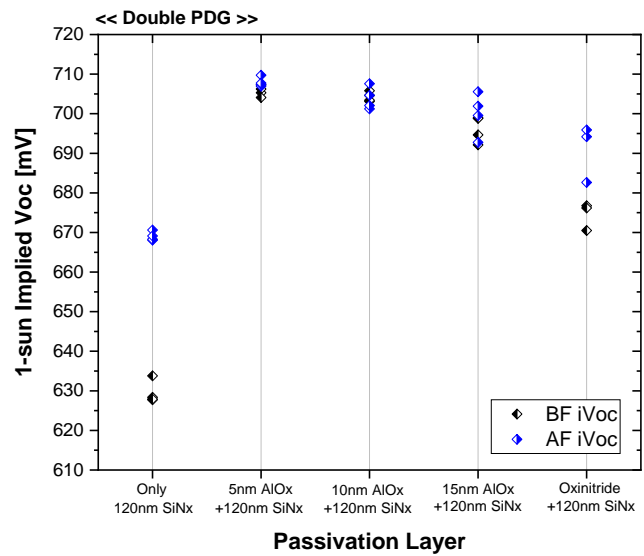

(c)

Figure 6. a) Carrier lifetime values for different passivation structures of samples after two PDG processes. b) PL-based carrier lifetime map of a representative sample after two PDG, passivated with a stack of 5nm AlOx/120nm SiNx, showing an average lifetime of 325.62 µs and local maxima as high as 790 µs. c) iVoc values for different passivation structures of samples after two PDG processes.

Our results demonstrate that best carrier lifetime values in multicrystalline p-type UMG-Si wafers can be achieved by a proper combination of P-diffusion gettering processing and passivation schemes based on $Al_2O_3/SiN_x$ layer stacks. The alumina thickness can be used as a free parameter for the joint optimization of two passivating contributions, stemming, respectively, from localized, fixed-charge induced field-effect at the Si/alumina interface, and from hydrogen in-diffusion across the alumina originating from the hydrogenated coatings. An optimum 5 nm interlayer alumina thickness, in combination with a low-temperature PDG pre-treatment, ensures bulk carrier lifetimes in UMG-Si wafers compatible with high-efficiency, industrially scalable PERC type cell production, insofar a good compromise between surface passivation quality, production speed and cost is met [26].

## 4. Conclusions

Effective carrier lifetime values up to 645 μs in multicrystalline p-type UMG-Silicon samples have been obtained as a result of the optimization of two consecutive Phosphorus Diffusion Gettering processes and iodine-ethanol reference surface passivation, with local values over 790 μs within individual grains. The overall improvement of the effective carrier lifetimes after PDG ranges between 20 and nearly 1000 times the starting values recorded from bare wafers, resulting in the highest values measured to date for this type of material. iVoc measurements reach up to 710 mV, opening the possibility for high efficiency results.

The positive response of the material to the sequential PDG treatments is uneven, with a major impact resulting from the first PDG carried out at a low temperature and a subsequent moderate but still significant contribution from the second PDG, that can be performed at high temperatures compatible with emitter formation. Indeed, the second PDG turns out to show ample tolerance with respect to processing parameter values, and carrier lifetime improvements are systematically observed after the second PDG under a relatively wide range of temperatures and times.

From the passivation studies, a compromise was found between the $Al_2O_3$ layer thickness and the provision of a hydrogenated $SiN_x$ capping layer, ensuring an effective passivation for our samples. 5 nm thick alumina assures good chemical passivation from the hydrogen contained in the 120 nm thick $SiN_x$ capping layer, adding up to the field-effect related passivating contribution stemming from localized interface charge at the alumina/Si interface. Our results also show that both contributions can be effectively preserved after firing steps related to subsequent metallization. Therefore, the electronic upgrading of UMG-Si wafers hereby proposed represents a good compromise between improved bulk properties, surface passivation efficacy, and manufacturing speed cost, paving the way for the utilisation of UMG-Si in the fabrication of high-efficiency solar cells.

**Acknowledgements**

This work is part of the R&D SOLAR-ERA.NET Cofund project CHEER-UP (PCI2019-111834-2), funded by MCIN/ AEI/10.13039/501100011033/ and the European Union, and also part of the Project MADRID-PV2 (S2018/EMT-4308) funded by the Regional Government of Madrid with the support from FEDER Funds, and Project 219M029 funded by TÜBİTAK. Juan José Torres, Manuel Funes and Hasan Asav are acknowledged for support in wafer processing and for fruitful discussions, and Ergi Dönerçark for his help in PL measurements. Aurinka PV is acknowledged for UMG wafer supply.

**References**

[1] United Nations Framework Convention on Climate Change, Paris Agreement, (2015).
[2] International Energy Agency, Key World Energy Statistics 2020, (2020). https://www.iea.org/reports/key-world-energy-statistics-2020.
[3] International Energy Agency, Database documentation, Electricity Information 2021 Edition, (2021).
[4] D. Ray, Lazard's Levelized Cost of Energy Analysis—Version 14.0, (2020) 21.
[5] IRENA Capacity Statistics 2021, (2021).
[6] M. Victoria, N. Haegel, I.M. Peters, R. Sinton, A. Jäger-Waldau, C. del Cañizo, C. Breyer, M. Stocks, A. Blakers, I. Kaizuka, K. Komoto, A. Smets, Solar photovoltaics

is ready to power a sustainable future, Joule. 5 (2021) 1041–1056. https://doi.org/10.1016/j.joule.2021.03.005.

[7] Fraunhofer ISE, Annual Photovoltaics Report, (2021).

[8] Bernreuter Research - Polysilicon Market Reports (from EnergyTrend), First Rays of Light at the End of the Polysilicon High-Price Tunnel. (2021). https://www.bernreuter.com/newsroom/polysilicon-news/article/first-rays-of-light-at-the-end-of-the-polysilicon-high-price-tunnel/.

[9] Bloomberg L.P., Silicon's 300% Surge Throws Another Price Shock at the World, (n.d.). https://www.bloomberg.com/news/articles/2021-10-01/silicon-s-300-surge-throws-another-price-shock-at-the-world.

[10] L. Méndez, E. Forniés, D. Garrain, A.P. Vázquez, A. Souto, Upgraded Metallurgical Grade Silicon for solar electricity production: a comparative Life Cycle Assessment, Science of the Total Environment. 789 (2021) 56. https://doi.org/10.1016/j.scitotenv.2021.147969.

[11] D. Macdonald, L.J. Geerligs, Recombination activity of interstitial iron and other transition metal point defects in p- and n-type crystalline silicon, Appl. Phys. Lett. 85 (2004) 4061–4063. https://doi.org/10.1063/1.1812833.

[12] S. Dubois, N. Enjalbert, J.P. Garandet, Effects of the compensation level on the carrier lifetime of crystalline silicon, Appl. Phys. Lett. 93 (2008) 032114. https://doi.org/10.1063/1.2961030.

[13] F. Rougieux, C. Samundsett, K.C. Fong, A. Fell, P. Zheng, D. Macdonald, J. Degoulange, R. Einhaus, M. Forster, High efficiency UMG silicon solar cells: impact of compensation on cell parameters, Progress in Photovoltaics: Research and Applications. 24 (2016) 725–734. https://doi.org/10.1002/pip.2729.

[14] A. Cuevas, M. Forster, F. Rougieux, D. Macdonald, Compensation Engineering for Silicon Solar Cells, Energy Procedia. 15 (2012) 67–77. https://doi.org/10.1016/j.egypro.2012.02.008.

[15] Forniés, Ceccaroli, Méndez, Souto, Pérez Vázquez, Vlasenko, Dieguez, Mass Production Test of Solar Cells and Modules Made of 100% UMG Silicon. 20.76% Record Efficiency, Energies. 12 (2019) 1495. https://doi.org/10.3390/en12081495.

[16] T.C. Thi, K. Koyama, K. Ohdaira, H. Matsumura, Effect of hydrogen on passivation quality of SiNx/Si-rich SiNx stacked layers deposited by catalytic chemical vapor deposition on c-Si wafers, Thin Solid Films. 575 (2015) 60–63. https://doi.org/10.1016/j.tsf.2014.10.016.

[17] E. Forniés, C. del Cañizo, L. Méndez, A. Souto, A. Pérez Vázquez, D. Garrain, UMG silicon for solar PV: from defects engineering to PV module degradation, Solar Energy. 220 (2021) 354–362. http://dx.doi.org/10.1016/j.solener.2021.03.076.

[18] D.B.M. Klaassen, A unified mobility model for device simulation—I. Model equations and concentration dependence, Solid-State Electronics. 35 (1992) 953–959. https://doi.org/10.1016/0038-1101(92)90325-7.

[19] F. Schindler, M. Forster, J. Broisch, J. Schön, J. Giesecke, S. Rein, W. Warta, M.C. Schubert, Towards a unified low-field model for carrier mobilities in crystalline silicon, Solar Energy Materials and Solar Cells. 131 (2014) 92–99. https://doi.org/10.1016/j.solmat.2014.05.047.

[20] N. Dasilva-Villanueva, S. Catalán-Gómez, D. Fuertes Marrón, C. Del Cañizo, New Insights into UMG Silicon Wafers Through Lifetime Spectroscopy, (2021).

[21] S. Catalán-Gómez, N. Dasilva-Villanueva, D. Fuertes Marrón, C. del Cañizo, Phosphorous Diffusion Gettering of Trapping Centers in Upgraded Metallurgical-Grade Solar Silicon, Phys. Status Solidi RRL. (2021) 2100054. https://doi.org/10.1002/pssr.202100054.

[22] J. Hofstetter, J.-F. Lelièvre, D.P. Fenning, M.I. Bertoni, T. Buonassisi, A. Luque, C. del Cañizo, Enhanced iron gettering by short, optimized low-temperature annealing after phosphorus emitter diffusion for industrial silicon solar cell processing, Physica Status Solidi c. 8 (2011) 759–762. https://doi.org/10.1002/pssc.201000334.

[23] H.T.T. Nguyen, N. Balaji, C. Park, N.M. Triet, A.H.T. Le, S. Lee, M. Jeon, D. Oh, V.A. Dao, J. Yi, $Al_2O_3$/SiON stack layers for effective surface passivation and anti-reflection of high efficiency n-type c-Si solar cells, Semicond. Sci. Technol. 32 (2017) 025005. https://doi.org/10.1088/1361-6641/32/2/025005.

[24] C. Kranz, J.H. Petermann, T. Dullweber, R. Brendel, Simulation-based Efficiency Gain Analysis of 21.2%-efficient Screen-printed PERC Solar Cells, Energy Procedia. 92 (2016) 109–115. https://doi.org/10.1016/j.egypro.2016.07.038.

[25] P. Preis, F. Buchholz, P. Diaz-Perez, J. Glatz-Reichenbach, C. Peter, S. Schmitt, J. Theobald, K. Peter, A.-K. Søiland, Towards 20% Solar Cell Efficiency Using Silicon from Metallurgical Process Route, Energy Procedia. 55 (2014) 589–595. https://doi.org/10.1016/j.egypro.2014.08.030.

[26] F. Souren, X. Gay, B. Dielissen, R. Görtzen, Upgrade of an Industrial Al-BSF Solar Cell Line Into PERC Using Spatial ALD Al2O3, in: 2016: pp. 946–950. https://doi.org/10.4229/EUPVSEC20162016-2BV.7.38.